\documentclass{revtex4}
\usepackage{amssymb}
\usepackage{amsmath}
\usepackage{graphicx}
\usepackage[font={footnotesize,it}]{caption}

\pdfoutput=1 
\input{tcilatex}

\begin{document}

\title{Electromagnetic wave propagation through inhomogeneous material layers%
}
\author{S. Habib Mazharimousavi}
\email{habib.mazhari@emu.edu.tr}
\author{Ashkan Roozbeh}
\email{ashkan.physics@gmail.com}
\author{M. Halilsoy}
\email{mustafa.halilsoy@emu.edu.tr}
\affiliation{Physics Department, Eastern Mediterranean University, G. Magusa north
Cyprus, Mersin 10 Turkey}

\begin{abstract}
We use Maxwell's equations in a sourceless, inhomogeneous medium with
continuous permeability $\mu \left( \mathbf{r}\right) $ and permittivity $%
\epsilon \left( \mathbf{r}\right) $ to study the wave propagation. The
general form of the wave equation is derived and by virtue of some physical
assumptions, including $\mu $ and $\epsilon $ as functions of $z,$ the
equation has been simplified. Finally by introducing a smooth step
dielectric variable we solve the wave equation in the corresponding medium
which is in conform with the well known results. Exact double-layer solution
in analytic form has been given in terms of the Heun functions.
\end{abstract}

\maketitle

\section{Introduction}

Having the electric permittivity of a system to be an isotropic and
continuous spatial function has numerous applications in chemistry,
biophysics and electronics \cite{1,2,3,4,5,6,7}. As an example, in heavy
doped regions the dielectric constant changes with the density of impurity
and so with the position. Such a region can be found in bipolar transistors, 
$p-n$ junctions and solar cells \cite{2}. It is also common that
permittivity is assumed to be an isotropic spatial function that changes
continuously in solvent region that allows us to formulate a computational
scheme \cite{3}. Other inhomogeneous medium can be seen in a biological
membrane, like a lipid bilayer surrounded by water. For this system
permittivity changes from a large value in the surrounding water to a lower
value in the bilayer \cite{4}. Also in electric double layers (EDL) due to
ion accumulation and hydration in the region we face with permittivity
variation, with effects on electric potential and interaction pressure
between surfaces \cite{5}. The main concern in such studies is the behavior
of a non-uniform mixed medium (a chemical solution or a $p-n$ junction) upon
an external static electric field using the Poisson-Boltzmann equation. This
differential equation (linear or non-linear form) can describe the
electrostatic effects extensively, ranging from a bimolecular system \cite{6}
to an electrolyte solution \cite{7}. The most regular form of this equation
can be written as%
\begin{equation}
\mathbf{\nabla }\cdot \left[ \epsilon \left( \mathbf{r}\right) \mathbf{%
\nabla }\psi \left( \mathbf{r}\right) \right] =-\rho _{f}\left( \mathbf{r}%
\right) -\sum_{i}c_{i}^{\infty }z_{i}q\lambda \left( \mathbf{r}\right) \exp %
\left[ \frac{-z_{i}q\psi \left( \mathbf{r}\right) }{\kappa _{B}T}\right] .
\end{equation}%
In this equation $\epsilon \left( \mathbf{r}\right) $ is the variable
dielectric, $\psi \left( \mathbf{r}\right) $ is the electrostatic potential
and $\rho _{f}\left( \mathbf{r}\right) $ represents the charge density of
the medium. Further, $z_{i}$ and $c_{i}^{\infty }$ show the charge and the
concentration of ions, ${T}$ is the temperature, $\kappa _{B}$ is the
Boltzmann constant and $\lambda \left( \mathbf{r}\right) $ is a factor that
depends on the accessibility of a position to ions in the medium.

In the present work, we study the wave propagation inside inhomogeneous
materials with the electric permittivity $\epsilon \left( \mathbf{r}\right) $
and the magnetic permeability $\mu \left( \mathbf{r}\right) $ as some
isotropic and continuous spatial functions. Without external sources (i.e., $%
\rho _{free}=0,$ $\mathbf{J}_{free}=0$) the wave equation for the electric
component of the electromagnetic wave (EMW) propagating in the medium with $%
\epsilon =\epsilon \left( \mathbf{r}\right) $ and $\mu =\mu \left( \mathbf{r}%
\right) $ is found to be%
\begin{equation}
\mathbf{\nabla }^{2}\mathbf{E}-\epsilon \mu \frac{\partial ^{2}\mathbf{E}}{%
\partial t^{2}}=-\left( \mathbf{\nabla }\tilde{\epsilon}.\mathbf{\nabla }%
\right) \mathbf{E}-\left( \mathbf{E}.\mathbf{\nabla }\right) \mathbf{\nabla }%
\tilde{\epsilon}-\mathbf{\nabla }\left( \tilde{\epsilon}+\tilde{\mu}\right)
\times \left( \mathbf{\nabla }\times \mathbf{E}\right)
\end{equation}%
in which $\tilde{\epsilon}=\ln \epsilon $ and $\tilde{\mu}=\ln \mu .$ A
similar equation can be written for magnetic component of the EMW but we
shall find $\mathbf{B}$ using one of the Maxwell's equation. A general
approach toward the solution for Eq. (2) may not be possible due to the
complicated form of the right hand side of the equation but by some
simplifications we shall find exact analytical solution for this wave-type
equation. There are some attempts to consider propagation of the
electromagnetic waves in the context of the general equation given in (2).
For the case that $\epsilon $ and $\mu $ are constant our equation reduces
to the standard wave-equation which can be found in every electromagnetic
book. R Diamant et al have considered this problem in number of papers \cite%
{8,9,10,11,12}. In their papers, R Diamant et al assume that $\mu =\mu _{0}$
and the results are mainly numeric. Although having numerical solution has
the worth in its own right, herein we are interested to obtain exact
analytical solutions for the tangent hyperbolic potential in the azimuthal
direction. For single layer our solution is expressed in hypergeometric
functions while the double layer case in terms of the rare Heun functions.
It is our belief that taking different profiles yield also exact solutions
expressible in terms of known mathematical functions. In obtaining exact
solutions we abide by the constraint condition $\mu =\mu _{0}=$constant,
through our inhomogenous layers. We shall assume further that the $\epsilon $
and $\mu $ vary only in one direction which is also the direction of
propagation. This direction throughout the paper will be $z-$direction.

Organization of the paper is as follows. In Sec. II we find the form of the
wave equation in a general system of coordinates and through some
specifications we simplify the wave equation. In Sec. III we introduce the
smooth step dielectric constant and solve the wave equation accordingly.
Sec. IV solves the problem of smooth double layers. Our conclusion is
presented in Sec. V.

\section{The Wave Equation}

\ \ We start with the sourceless $(\rho _{free}=0,\mathbf{J}_{free}=0)$
Maxwell's equations in a medium with variable permittivity $\epsilon $ and
permeability $\mu $ which are given by 
\begin{equation}
\mathbf{\nabla .B}=\mathbf{0,}\text{ \ \ }\mathbf{\nabla .D}=\mathbf{0,}%
\text{ \ \ }\mathbf{\nabla \times E}=\mathbf{-}\frac{\partial \mathbf{B}}{%
\partial t}\text{ \ and }\mathbf{\nabla \times H}=\frac{\partial \mathbf{D}}{%
\partial t}.
\end{equation}%
\newline
Herein $\mathbf{D=}\epsilon \mathbf{E}$ is the displacement vector, $\mathbf{%
E}$ is the electric field, $\mathbf{H=}\frac{1}{\mu }\mathbf{B}$ is the
auxiliary magnetic field and $\mathbf{B}$ is the magnetic field. To obtain
Eq. (2) from the Maxwell's equations (3) we start from Faraday's law and we
act $\mathbf{\nabla \times }$ from the left i.e., $\mathbf{\nabla \times }%
\left( \mathbf{\nabla \times E}\right) =\mathbf{-}\frac{\partial \mathbf{%
\nabla \times B}}{\partial t}$ which yields%
\begin{eqnarray}
\mathbf{\nabla }\left( \mathbf{\nabla .E}\right) -\mathbf{\nabla }^{2}%
\mathbf{E} &\mathbf{=}&\mathbf{-}\frac{\partial }{\partial t}\mathbf{\nabla
\times }\left( \mu \mathbf{H}\right) \\
&=&\mathbf{-}\frac{\partial }{\partial t}\left[ \mu \mathbf{\nabla \times
H-H\times \nabla }\mu \right] ,  \notag
\end{eqnarray}%
and equivalently%
\begin{eqnarray}
\mathbf{\nabla }^{2}\mathbf{E-\nabla }\left( \mathbf{\nabla .E}\right) &=&%
\frac{\partial }{\partial t}\left[ \mu \frac{\partial \mathbf{D}}{\partial t}%
\mathbf{-B\times }\frac{\mathbf{\nabla }\mu }{\mu }\right] \\
&=&\mu \epsilon \frac{\partial ^{2}\mathbf{E}}{\partial t^{2}}\mathbf{-}%
\frac{\partial \mathbf{B}}{\partial t}\mathbf{\times \nabla }\ln \mu . 
\notag
\end{eqnarray}%
Next we rearrange the terms to get%
\begin{eqnarray}
\mathbf{\nabla }^{2}\mathbf{E-}\mu \epsilon \frac{\partial ^{2}\mathbf{E}}{%
\partial t^{2}} &=&\mathbf{\nabla }\left( \mathbf{\nabla .E}\right) \mathbf{-%
}\frac{\partial \mathbf{B}}{\partial t}\mathbf{\times \nabla }\ln \mu \\
&=&\mathbf{\nabla }\left( \mathbf{\nabla .}\left( \frac{\mathbf{D}}{\epsilon 
}\right) \right) \mathbf{+}\left( \mathbf{\nabla \times E}\right) \mathbf{%
\times \nabla }\tilde{\mu}  \notag \\
&=&-\mathbf{\nabla }\left( \mathbf{D.}\frac{\mathbf{\nabla }\epsilon }{%
\epsilon ^{2}}\right) \mathbf{+}\left( \mathbf{\nabla \times E}\right) 
\mathbf{\times \nabla }\tilde{\mu}.  \notag
\end{eqnarray}%
Finally one finds%
\begin{eqnarray}
\mathbf{\nabla }^{2}\mathbf{E-}\mu \epsilon \frac{\partial ^{2}\mathbf{E}}{%
\partial t^{2}} &=&-\mathbf{\nabla }\left( \mathbf{E.\nabla }\tilde{\epsilon}%
\right) \mathbf{-\nabla }\tilde{\mu}\times \left( \mathbf{\nabla \times E}%
\right) \\
&=&-\left[ \left( \mathbf{\nabla }\tilde{\epsilon}.\mathbf{\nabla }\right) 
\mathbf{E+}\left( \mathbf{E.\nabla }\right) \mathbf{\nabla }\tilde{\epsilon}+%
\mathbf{\nabla }\tilde{\epsilon}\times \left( \mathbf{\nabla \times E}%
\right) +\mathbf{E\times }\left( \mathbf{\nabla \times \nabla }\tilde{%
\epsilon}\right) \right] \mathbf{-\nabla }\tilde{\mu}\times \left( \mathbf{%
\nabla \times E}\right)  \notag \\
&=&-\left( \mathbf{\nabla }\tilde{\epsilon}.\mathbf{\nabla }\right) \mathbf{%
E-}\left( \mathbf{E.\nabla }\right) \mathbf{\nabla }\tilde{\epsilon}-\mathbf{%
\nabla }\left( \tilde{\epsilon}+\tilde{\mu}\right) \times \left( \mathbf{%
\nabla \times E}\right) ,  \notag
\end{eqnarray}%
which is nothing but Eq. (2). We comment that in finding Eq. (7) we have
used the mathematical formula: $\mathbf{\nabla \times }\left( \mathbf{\nabla
\times A}\right) =\mathbf{\nabla }\left( \mathbf{\nabla .A}\right) -\mathbf{%
\nabla }^{2}\mathbf{A}$ and $\mathbf{\nabla }\left( \mathbf{A.B}\right)
=\left( \mathbf{B.\nabla }\right) \mathbf{A+}\left( \mathbf{A.\nabla }%
\right) \mathbf{B+B\times }\left( \mathbf{\nabla \times A}\right) \mathbf{%
+A\times }\left( \mathbf{\nabla \times B}\right) $ and $\tilde{\epsilon}=\ln
\epsilon $ and $\tilde{\mu}=\ln \mu .$ Upon taking into account that in our
study $\epsilon =\epsilon \left( z\right) $ and $\mu =\mu \left( z\right) $
the general wave equation (2) becomes \newline
\begin{equation}
\mathbf{\nabla }^{2}\mathbf{E}-\epsilon \mu \frac{\partial ^{2}\mathbf{E}}{%
\partial t^{2}}=-\left( \tilde{\epsilon}^{\prime }\frac{\partial }{\partial z%
}\right) \mathbf{E-}\left( \mathbf{E}.\mathbf{\nabla }\right) \tilde{\epsilon%
}^{\prime }\hat{k}-\left( \tilde{\epsilon}^{\prime }+\tilde{\mu}^{\prime
}\right) \hat{k}\times \left( \mathbf{\nabla }\times \mathbf{E}\right) ,
\end{equation}%
in which a prime $^{\prime }$ denotes $\frac{d}{dz}$\noindent . Our further
simplification is to consider that the electric and magnetic components of
the EMW vary in $z$ direction only. This assumption is due to the symmetry
of the medium which is physically acceptable. Our latter equation then
reduces effectively into the following $1-$dimensional three equations for
the electric components%
\begin{equation}
\left( \frac{\mathbf{\partial }^{2}}{\partial z^{2}}-\mu \epsilon \frac{%
\partial ^{2}}{\partial t^{2}}\right) E_{x}=\tilde{\mu}^{\prime }\frac{%
\partial E_{x}}{\partial z},
\end{equation}%
\newline
\begin{equation}
\left( \frac{\mathbf{\partial }^{2}}{\partial z^{2}}-\mu \epsilon \frac{%
\partial ^{2}}{\partial t^{2}}\right) E_{y}=\tilde{\mu}^{\prime }\frac{%
\partial E_{y}}{\partial z},
\end{equation}%
\newline
\begin{equation}
\left( \frac{\mathbf{\partial }^{2}}{\partial z^{2}}-\mu \epsilon \frac{%
\partial ^{2}}{\partial t^{2}}\right) E_{z}=-\tilde{\epsilon}^{\prime
}E_{z}^{\prime }-E_{z}\tilde{\epsilon}^{\prime \prime }.
\end{equation}%
\newline
Next we consider \newline
\begin{equation}
\mathbf{E}\left( \mathbf{r,}t\right) \mathbf{=\bar{E}}\left( z\right)
e^{i\omega t}
\end{equation}%
\newline
in which $\omega $ is the angular frequency of the wave. A substitution in
(9) and (10) yields:\newline
\begin{equation}
\left( \frac{d^{2}}{dz^{2}}+\mu \epsilon \omega ^{2}\right) \bar{E}%
_{i}\left( z\right) =\frac{\mu ^{\prime }}{\mu }\frac{d\bar{E}_{i}\left(
z\right) }{dz}
\end{equation}%
\newline
for $i=x,y.$ To have the third equation (i.e. Eq. (11)) satisfied we set $%
E_{z}=0$ (i.e. no longitudinal component) which is equivalent with the
propagation of the wave in $z-$direction. Having symmetry with respect to $x$
and $y$ one can always rotate the system of coordinates in $z$ direction
such that $\mathbf{\bar{E}}\left( z\right) $ aligns with one of the
coordinates (say $x$). Hence we are left with only one equation in $x-$%
direction which is expressed as%
\begin{equation}
\left( \frac{d^{2}}{dz^{2}}+\mu \epsilon \omega ^{2}\right) \bar{E}%
_{x}\left( z\right) =\frac{\mu ^{\prime }}{\mu }\frac{d\bar{E}_{x}\left(
z\right) }{dz}
\end{equation}%
and the other two components of the electric field are zero i.e., $\mathbf{E}%
\left( \mathbf{r,}t\right) \mathbf{=\hat{x}}\bar{E}_{x}\left( z\right)
e^{i\omega t}.$ To do further investigation one must know the form of $\mu $
and $\epsilon $ in terms of $z.$ For instance, with $\mu =\mu _{0}=cont.$
and $\epsilon =\epsilon _{0}=cont.$ one finds%
\begin{equation}
\left( \frac{d^{2}}{dz^{2}}+\frac{\omega ^{2}}{c^{2}}\right) \bar{E}_{x}=0
\end{equation}%
\newline
which admits 
\begin{eqnarray}
\bar{E}_{x} &=&\bar{E}_{0x}e^{\mp ikz} \\
\text{(}\bar{E}_{0x} &=&\text{constant)}  \notag
\end{eqnarray}%
\newline
where $k=$ $\frac{\omega }{c}$ and the plane wave is propagating in $\pm z$
direction. Our final remark in this section is on the form of the wave
equation: if one considers $\mu ^{\prime }=0$ it reduces to the form of the
standard wave equation, but owing to the form of $\epsilon =\epsilon \left(
z\right) $ its solution differs from the standard wave equation.

\section{Smooth step dielectric constant}

As an example let's consider $\mu =K_{m}\mu _{0}=cons.$ and $\epsilon
=K_{e}\left( z\right) \epsilon _{0},$ where $K_{e}\left( z\right) $ stands
for a smooth function of $z$ given by \cite{11} 
\begin{equation}
K_{e}\left( z\right) =K_{2}-\frac{\Delta K}{4}\left( 1-\tanh \left(
az\right) \right) ^{2},\text{ }
\end{equation}%
in which $a$ is a positive real constant, $\Delta K=K_{2}-K_{1},$ $%
K_{2}=\lim_{z\rightarrow \infty }K_{e}\left( z\right) $ and $%
K_{1}=\lim_{z\rightarrow -\infty }K_{e}\left( z\right) .$ The wave equation
(13) becomes%
\begin{equation}
\left( \frac{d^{2}}{dz^{2}}+\frac{\omega ^{2}}{c^{2}}K_{m}K_{e}\left(
z\right) \right) \bar{E}_{x}\left( z\right) =0,
\end{equation}%
\newline
which after defining the following new parameters 
\begin{equation}
\kappa ^{2}=\frac{\omega ^{2}}{c^{2}}K_{m}K_{1},\text{ }\nu ^{2}=\frac{%
\omega ^{2}}{c^{2}}K_{m}K_{2}
\end{equation}%
and considering a new variable 
\begin{equation}
\xi =-e^{-2az}
\end{equation}%
together with a redefinition of the electric field%
\begin{equation}
\bar{E}_{x}\left( z\right) =\left( -\xi \right) ^{-i\nu }F\left( \xi \right)
,
\end{equation}%
it turns (with $^{\prime }=\frac{d}{d\xi }$) into 
\begin{equation}
\xi F^{\prime \prime }+\left( 1-2i\nu \right) F^{\prime }+\frac{1}{4a^{2}}%
\left[ \frac{\left( 1-4a^{2}\right) \nu ^{2}}{\xi }+\frac{\nu ^{2}-\kappa
^{2}}{1-\xi }-\frac{\nu ^{2}-\kappa ^{2}}{\left( 1-\xi \right) ^{2}}\right]
F=0.
\end{equation}%
Having singularities at $\xi =0$ and $\xi =1$ suggests to replace further%
\begin{equation}
F\left( \xi \right) =\xi ^{\sigma }\left( \xi -1\right) ^{\rho }G\left( \xi
\right)
\end{equation}%
which after some manipulation and choices 
\begin{equation}
\sigma =\frac{i\nu \left( 2a-1\right) }{2a}
\end{equation}%
and%
\begin{equation}
\rho =\frac{1}{2}\left( 1-\frac{1}{a}\sqrt{a^{2}+\nu ^{2}-\kappa ^{2}}\right)
\end{equation}%
it reduces to the following Hypergeometric differential equation (HDE) \cite%
{13}%
\begin{equation}
\xi \left( \xi -1\right) G^{\prime \prime }+\left[ \frac{i\nu -a}{a}-\left( 
\frac{i\nu -2a}{a}+\frac{1}{a}\sqrt{a^{2}+\nu ^{2}-\kappa ^{2}}\right) \xi %
\right] G^{\prime }-\frac{i\nu -a}{2a^{2}}\left( a-\sqrt{a^{2}+\nu
^{2}-\kappa ^{2}}\right) G=0.
\end{equation}%
Comparing with the standard form of the Hypergeometric DE 
\begin{equation}
\xi \left( \xi -1\right) G^{\prime \prime }+\left[ \left( \alpha +\beta
+1\right) \xi -\gamma \right] G^{\prime }+\alpha \beta G=0,
\end{equation}%
one finds%
\begin{equation}
\alpha =\frac{1}{2a}\left[ a-\sqrt{a^{2}+\nu ^{2}-\kappa ^{2}}-i\left( \nu
+\kappa \right) \right] ,
\end{equation}%
\begin{equation}
\beta =\frac{1}{2a}\left[ a-\sqrt{a^{2}+\nu ^{2}-\kappa ^{2}}-i\left( \nu
-\kappa \right) \right] ,
\end{equation}%
and%
\begin{equation}
\gamma =\frac{a-i\nu }{a}.
\end{equation}%
The general solution for the above HDE can be written as%
\begin{equation}
G=C_{1}F\left( \alpha ,\beta ;\gamma ;\xi \right) +C_{2}\xi ^{1-\gamma
}F\left( \alpha -\gamma +1,\beta -\gamma +1;2-\gamma ;\xi \right)
\end{equation}%
in which $C_{1}$ and $C_{2}$ are two integration constants. Going to the
original variables now, the general solution for the Electric field is given
by%
\begin{eqnarray}
\bar{E}_{x}\left( z\right) &=&\tilde{C}_{1}\left( -\xi \right) ^{-\frac{i\nu 
}{2a}}\left( 1-\xi \right) ^{\rho }F\left( \alpha ,\beta ;\gamma ;\xi
\right) + \\
&&\tilde{C}_{2}\left( -\xi \right) ^{\frac{i\nu }{2a}}\left( 1-\xi \right)
^{\rho }F\left( \alpha -\gamma +1,\beta -\gamma +1;2-\gamma ;\xi \right) , 
\notag
\end{eqnarray}%
where $\tilde{C}_{1}=C_{1}\left( -1\right) ^{\sigma +\rho }$ and $\tilde{C}%
_{2}=C_{2}\left( -1\right) ^{\sigma +\rho +1-\gamma }.$ Let us consider that
the wave comes from $z=-\infty $ and goes toward $z=+\infty .$ Also we
recall that $\lim_{z\rightarrow \infty }K_{e}\left( z\right) =K_{2}=cons.$
which implies that $\lim_{z\rightarrow \infty }\bar{E}_{x}\left( z\right)
\sim e^{i\frac{\omega }{c}\sqrt{K_{m}K_{2}}z}=e^{i\nu z}.$ Once $%
z\rightarrow +\infty $ it is clear that $\xi =-e^{-2az}\rightarrow 0$, and
upon knowing that $F\left( \alpha ,\beta ;\gamma ;0\right) =1,$ makes the
limit of the electric field to be%
\begin{equation}
\lim_{z\rightarrow \infty }\bar{E}_{x}\left( z\right) =\tilde{C}_{1}e^{i\nu
z}+\tilde{C}_{2}e^{-i\nu z}.
\end{equation}%
This suggests that for this choice we must set $\tilde{C}_{2}=0,$ which
casts the solution into%
\begin{equation}
\bar{E}_{x}\left( z\right) =\tilde{C}_{1}\left( -\xi \right) ^{-\frac{i\nu }{%
2a}}\left( 1-\xi \right) ^{\rho }F\left( \alpha ,\beta ;\gamma ;\xi \right) ,
\end{equation}%
so that%
\begin{equation}
\lim_{z\rightarrow \infty }\bar{E}_{x}\left( z\right) =\tilde{C}_{1}e^{i\nu
z}=E_{02}e^{i\nu z}.
\end{equation}%
Herein we consider the amplitude of the transmitted wave as $E_{02}.$

The limit of $z\rightarrow -\infty $ (and consequently $\xi \rightarrow
-\infty $) can be found once we apply the following properties for the
Hypergeometric functions \cite{14}%
\begin{eqnarray}
F\left( \alpha ,\beta ;\gamma ;\xi \right) &=&\frac{\Gamma \left( \gamma
\right) \Gamma \left( \beta -\alpha \right) }{\Gamma \left( \beta \right)
\Gamma \left( \gamma -\alpha \right) }\left( -1\right) ^{\alpha }\xi
^{-\alpha }F\left( \alpha ,\alpha +1-\gamma ;\alpha +1-\beta ;\frac{1}{\xi }%
\right) + \\
&&\frac{\Gamma \left( \gamma \right) \Gamma \left( \alpha -\beta \right) }{%
\Gamma \left( \alpha \right) \Gamma \left( \gamma -\alpha \right) }\left(
-1\right) ^{\beta }\xi ^{-\beta }F\left( \beta ,\beta +1-\gamma ;\beta
+1-\alpha ;\frac{1}{\xi }\right) .  \notag
\end{eqnarray}%
Hence 
\begin{equation}
\lim_{\substack{ z\rightarrow -\infty  \\ \xi \rightarrow -\infty }}F\left(
\alpha ,\beta ;\gamma ;\xi \right) =\lim_{\substack{ z\rightarrow -\infty 
\\ \xi \rightarrow -\infty }}\frac{\Gamma \left( \gamma \right) \Gamma
\left( \beta -\alpha \right) }{\Gamma \left( \beta \right) \Gamma \left(
\gamma -\alpha \right) }\left( -1\right) ^{\alpha }\xi ^{-\alpha }+\frac{%
\Gamma \left( \gamma \right) \Gamma \left( \alpha -\beta \right) }{\Gamma
\left( \alpha \right) \Gamma \left( \gamma -\alpha \right) }\left( -1\right)
^{\beta }\xi ^{-\beta },
\end{equation}%
which upon (32), one finds%
\begin{equation}
\lim_{\substack{ z\rightarrow -\infty  \\ \xi \rightarrow -\infty }}\bar{E}%
_{x}\left( z\right) =\tilde{C}_{1}\left( \frac{\Gamma \left( \gamma \right)
\Gamma \left( \beta -\alpha \right) }{\Gamma \left( \beta \right) \Gamma
\left( \gamma -\alpha \right) }\left( -\xi \right) ^{-\alpha -\frac{i\nu }{2a%
}+\rho }+\frac{\Gamma \left( \gamma \right) \Gamma \left( \alpha -\beta
\right) }{\Gamma \left( \alpha \right) \Gamma \left( \gamma -\alpha \right) }%
\left( -\xi \right) ^{-\beta -\frac{i\nu }{2a}+\rho }\right) .
\end{equation}%
Now one can show that%
\begin{eqnarray}
-\alpha -\frac{i\nu }{2a}+\rho &=&+i\frac{\kappa }{2a}, \\
-\beta -\frac{i\nu }{2a}+\rho &=&-i\frac{\kappa }{2a}
\end{eqnarray}%
which finally yields%
\begin{equation}
\lim_{\substack{ z\rightarrow -\infty  \\ \xi \rightarrow -\infty }}\bar{E}%
_{x}\left( z\right) =\tilde{C}_{1}\left( \frac{\Gamma \left( \gamma \right)
\Gamma \left( \beta -\alpha \right) }{\Gamma \left( \beta \right) \Gamma
\left( \gamma -\alpha \right) }e^{-i\mu z}+\frac{\Gamma \left( \gamma
\right) \Gamma \left( \alpha -\beta \right) }{\Gamma \left( \alpha \right)
\Gamma \left( \gamma -\alpha \right) }e^{i\mu z}\right) =E_{01}^{\prime
}e^{-i\mu z}+E_{01}e^{i\mu z}.
\end{equation}%
Note that 
\begin{equation}
E_{01}=\frac{\Gamma \left( \gamma \right) \Gamma \left( \alpha -\beta
\right) }{\Gamma \left( \alpha \right) \Gamma \left( \gamma -\alpha \right) }%
E_{02}
\end{equation}%
is the amplitude of the transmitted wave for $z\rightarrow \infty $ and 
\begin{equation}
E_{01}^{\prime }=\frac{\Gamma \left( \gamma \right) \Gamma \left( \beta
-\alpha \right) }{\Gamma \left( \beta \right) \Gamma \left( \gamma -\alpha
\right) }E_{02}
\end{equation}%
is the amplitude of the reflected wave for $z\rightarrow -\infty .$ Once
more we comment that $E_{02},$ is the amplitude of the transmitted wave
while $E_{01}$ and $E_{01}^{\prime }$ are the amplitudes of the incident and
reflected waves, respectively. Having these definitions together with the
above equations one obtains the reflection and transmission coefficients of
the wave as%
\begin{equation}
R=\frac{E_{01}^{\prime }}{E_{01}}=\frac{\Gamma \left( \alpha \right) \Gamma
\left( \beta -\alpha \right) }{\Gamma \left( \beta \right) \Gamma \left(
\alpha -\beta \right) }
\end{equation}%
and%
\begin{equation}
T=\frac{E_{02}}{E_{01}}=\frac{\Gamma \left( \alpha \right) \Gamma \left(
\gamma -\alpha \right) }{\Gamma \left( \gamma \right) \Gamma \left( \alpha
-\beta \right) }.
\end{equation}%
As one observes $R$ and $T$ depend on $\alpha ,\beta $ and $\gamma $ which
are complex parameters. This means that $R$ and $T$ are complex numbers too.
In a particular case in which $\alpha ,\beta $ and $\gamma $ are given one
can calculate the value of the Gamma functions and accordingly $R$ and $T$
are obtained. Irrespective to the value of the physical parameters, there
exist always a complex number assigned to $R$ and $T.$ In the limit of a
sharp step dielectric i.e., $a\rightarrow \infty $ one finds%
\begin{equation}
\lim_{a\rightarrow \infty }R=\lim_{a\rightarrow \infty }\frac{\Gamma \left(
\alpha \right) \Gamma \left( \beta -\alpha \right) }{\Gamma \left( \beta
\right) \Gamma \left( \alpha -\beta \right) }=\frac{\kappa -\nu }{\kappa
+\nu }=\frac{\sqrt{K_{m}K_{1}}-\sqrt{K_{m}K_{2}}}{\sqrt{K_{m}K_{1}}+\sqrt{%
K_{m}K_{2}}}=\frac{n_{1}-n_{2}}{n_{1}+n_{2}}
\end{equation}%
\begin{equation}
\lim_{a\rightarrow \infty }T=\lim_{a\rightarrow \infty }\frac{\Gamma \left(
\gamma \right) \Gamma \left( \alpha -\beta \right) }{\Gamma \left( \alpha
\right) \Gamma \left( \gamma -\alpha \right) }=\frac{2\kappa }{\kappa +\nu }=%
\frac{2\sqrt{K_{m}K_{1}}}{\sqrt{K_{m}K_{1}}+\sqrt{K_{m}K_{2}}}=\frac{2n_{1}}{%
n_{1}+n_{2}}
\end{equation}%
where $n_{1}$ and $n_{2}$ are the optical indices of the spaces at $z<0$ and 
$z>0,$ respectively \cite{15}. In Fig. 1 we plot $\func{Re}\left( \bar{E}%
_{x}\left( z\right) \right) $ in terms of $z$ together with $K_{e}\left(
z\right) $ for certain values of parameters. The smooth change in the
dielectric constant and amplitude of the electric field are clear.

\begin{figure}[tbp]
\includegraphics{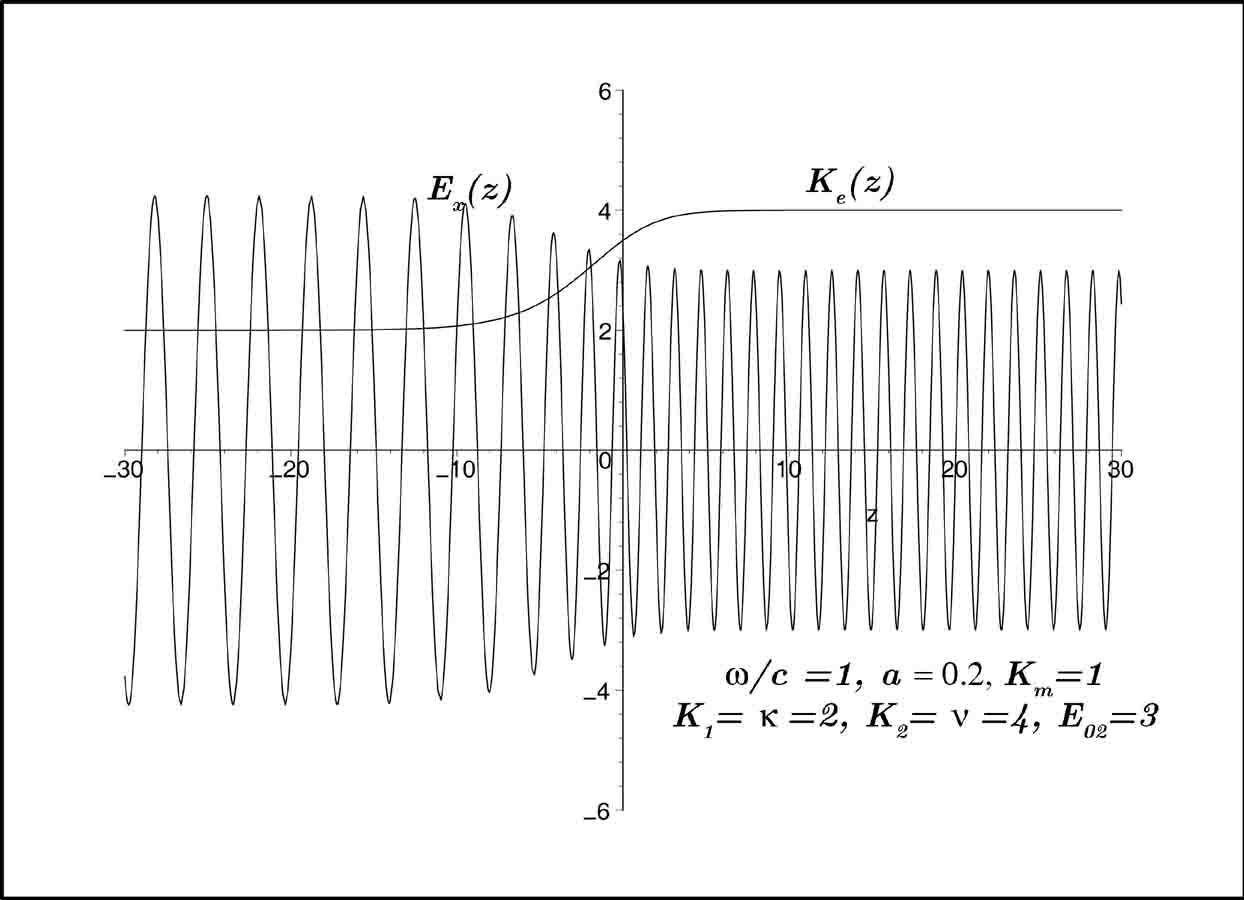}
\caption{A plot of $\func{Re}\left( \bar{E}_{x}\left( z\right) \right) $ in
terms of $z$ together with $K_{e}\left( z\right) $ for certain values of the
parameters. The smooth changes in the dielectric constant and amplitude of
the electric field are clear.}
\end{figure}

To complete our solution we rewrite the electric field of a plane wave
solution which propagates in positive $z$ direction,%
\begin{equation}
\mathbf{E}\left( \mathbf{r,}t\right) \mathbf{=\hat{x}}E_{02}\left( -\xi
\right) ^{-\frac{i\nu }{2a}}\left( 1-\xi \right) ^{\rho }F\left( \alpha
,\beta ;\gamma ;\xi \right) e^{i\omega t},
\end{equation}%
together with the magnetic field which can be found by applying the same
Maxwell's equation as we started with i.e., $\mathbf{\nabla \times E}=%
\mathbf{-}\frac{\partial \mathbf{B}}{\partial t}.$ Furthermore we need also
to use \cite{16} 
\begin{equation}
\frac{\partial }{\partial z}F\left( \alpha ,\beta ;\gamma ;\xi \right) =%
\frac{\partial \xi }{\partial z}\frac{\partial }{\partial z}F\left( \alpha
,\beta ;\gamma ;\xi \right) =2a\xi \frac{\alpha \beta }{\gamma }F\left(
\alpha +1,\beta +1;\gamma +1;\xi \right)
\end{equation}%
and in general if we take $\mathbf{E}\left( \mathbf{r,}t\right) \mathbf{=%
\hat{x}}E\left( z\right) e^{i\omega t}$ then $\mathbf{B}\left( \mathbf{r,}%
t\right) =-i\omega \mathbf{\hat{y}}\frac{\partial E\left( z\right) }{%
\partial z}e^{i\omega t}$ or finally 
\begin{equation}
\mathbf{B}\left( \mathbf{r,}t\right) \mathbf{=\hat{y}}\frac{iE_{02}}{2\omega
a}\left( -\xi \right) ^{-\frac{i\nu }{2a}}\left( 1-\xi \right) ^{\rho }\left[
\frac{2a\alpha \beta }{\gamma }F\left( \beta +1,\alpha +1;\gamma +1;\xi
\right) +\left( \frac{\nu i}{\xi }-\frac{2a\rho }{\left( 1-\xi \right) }%
\right) F\left( \alpha ,\beta ;\gamma ;\xi \right) \right] e^{i\omega t}.
\end{equation}

\section{Smooth double layers}

Another application of the general equation is to find the EMW passing
through a double layer thick shell. For this let us consider $K_{m}=const.$
with%
\begin{equation}
K_{e}\left( z\right) =K_{1}+\frac{K_{2}-K_{1}}{2}\left( \tanh az-\tanh
a\left( z-L\right) \right) .
\end{equation}%
Here $a$ is a constant which in the limit $a\rightarrow \infty ,$ $L$
becomes the thickness of a flat double layer dielectric of dielectric
constant $K_{2}$ located inside another medium of dielectric constant $%
K_{1}. $ Next, we rewrite the wave equation (13) in this matter: 
\begin{equation}
\left( \frac{d^{2}}{dz^{2}}+\left\{ \kappa ^{2}+\frac{\nu ^{2}-\kappa ^{2}}{2%
}\left( \tanh az-\tanh a\left( z-L\right) \right) \right\} \right) \bar{E}%
_{x}\left( z\right) =0,
\end{equation}%
in which $\kappa $ and $\nu $ are defined in (19). We follow a similar
change of variable given by (21) which modifies the latter equation as%
\begin{equation}
\xi ^{2}E^{\prime \prime }\left( \xi \right) +\xi E^{\prime }\left( \xi
\right) +\frac{1}{4a^{2}}\left( \kappa ^{2}-\frac{\left( \nu ^{2}-\kappa
^{2}\right) \left( \lambda -1\right) }{\lambda \left( \xi -1\right) \left(
\xi -\frac{1}{\lambda }\right) }\right) E\left( \xi \right) =0
\end{equation}%
in which $\lambda =\exp \left( 2aL\right) .$ Now, we replace $E\left( \xi
\right) =\xi ^{\sigma }H\left( \xi \right) $ with $\sigma =-i\kappa /2a$ to
find 
\begin{equation}
H^{\prime \prime }\left( \xi \right) +\frac{1-i\kappa /a}{\xi }H^{\prime
}\left( \xi \right) +\frac{\left( \kappa ^{2}-\nu ^{2}\right) \left( \lambda
-1\right) }{4a^{2}\lambda \xi \left( \xi -1\right) \left( \xi -\frac{1}{%
\lambda }\right) }H\left( \xi \right) =0
\end{equation}%
which is a Heun differential equation \cite{17} of the form%
\begin{equation}
w^{\prime \prime }\left( z\right) +\left( \frac{\gamma }{z}+\frac{\delta }{%
z-1}+\frac{\epsilon }{z-p}\right) w^{\prime }\left( z\right) +\frac{\alpha
\beta z-q}{z\left( z-1\right) \left( z-p\right) }w\left( z\right) =0
\end{equation}%
in which $\epsilon =\alpha +\beta -\gamma -\delta +1,$ with a general
solution%
\begin{equation}
w\left( z\right) =C_{1}HeunG\left( p,q,\alpha ,\beta ,\gamma ,\delta
,z\right) +C_{2}z^{1-\gamma }HeunG\left( p,q-\left( p\delta +\epsilon
\right) \left( \gamma -1\right) ,\beta -\gamma +1,\alpha -\gamma +1,2-\gamma
,\delta ,z\right) .
\end{equation}%
Herein $C_{1}$ and $C_{2}$ are two integration constants. \ Following the
general solution we find the final solution of the wave equation as%
\begin{eqnarray}
E\left( \xi \right) &=&C_{1}\left( -\xi \right) ^{\frac{-i\kappa }{2a}%
}HeunG\left( \frac{1}{\lambda e},\frac{\left( \nu ^{2}-\kappa ^{2}\right)
\left( \lambda -1\right) }{4a^{2}\lambda },0,\frac{-i\kappa }{a},\frac{%
a-i\kappa }{a},0,\xi \right) + \\
&&C_{2}\left( -\xi \right) ^{\frac{i\kappa }{2a}}HeunG\left( \frac{1}{%
\lambda },\frac{\left( \nu ^{2}-\kappa ^{2}\right) \left( \lambda -1\right) 
}{4a^{2}\lambda },0,\frac{i\kappa }{a},\frac{a+i\kappa }{a},0,\xi \right) . 
\notag
\end{eqnarray}%
Upon considering $HeunG\left( p,q,\alpha ,\beta ,\gamma ,\delta ,0\right)
=1, $ one easily finds that $\lim_{\substack{ z\rightarrow \infty  \\ \xi
\rightarrow 0}}E\left( \xi \right) =C_{1}\exp \left( i\kappa z\right)
+C_{2}\exp \left( -i\kappa z\right) $ which after assuming that the wave
starts from $z=-\infty $ and propagates toward $z=+\infty $ one must set $%
C_{2}=0$ and $C_{1}=E_{03}.$ These, therefore, imply%
\begin{equation}
E\left( \xi \right) =E_{03}\left( -\xi \right) ^{\frac{-i\mu }{2a}%
}HeunG\left( \frac{1}{\lambda },\frac{\left( \nu ^{2}-\kappa ^{2}\right)
\left( \lambda -1\right) }{4a^{2}\lambda },0,\frac{-i\mu }{a},\frac{%
a-i\kappa }{a},0,\xi \right) ,
\end{equation}%
where $E_{03}$ is the amplitude of the electric field at the limit $%
z\rightarrow +\infty $. Fig. 2 displays the continuous passage of EMW
through medium of the double-layers located at $z=0$ and $z=4.$

\begin{figure}[tbp]
\includegraphics{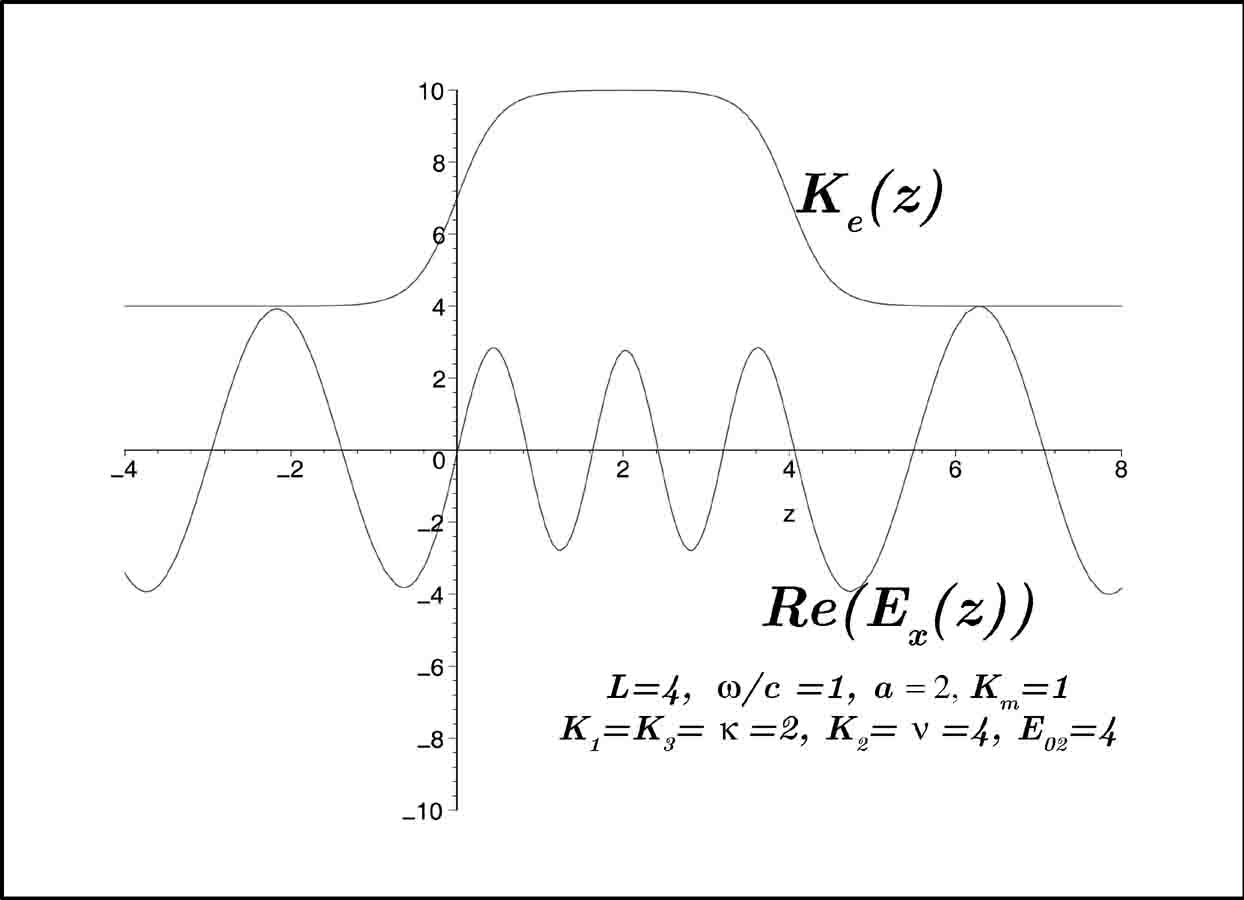}
\caption{The incoming EMW from $z<0$ encounters with the first layer at $z=0$%
. $Re(E_{x}\left( z\right) )$ has similar structure after crossing the
second layer at $z=4$. In between, $0<z<4,$ the oscillatory behaviour
evidently changes. }
\end{figure}

\section{\textbf{Conclusion}}

In conclusion we have considered media with variable permeability and
permittivity varying with $z-$direction, alone. The general form of the wave
equation in such a non-uniform medium is presented. Some simplifications
have been considered such as, $\epsilon $ and $\mu $ are only functions of $%
z $ which is also the direction of the propagation of the possible EMW. In
such a framework we have used the sourceless Maxwell's equations to find a
wave equation for the propagation of the EMW. We solved the wave equation
for a smooth-step, variable dielectric given the constant permeability. In
the limits we obtain the well known reflection and transmission coefficients
of the plane waves in normal incidence on the interface plane between two
dielectrics. Our results have been presented analytically and schematically.
Studying this interesting problem further with more layers is our future
plan. The applications of such extremal theories may not be so clear yet but
we believe that with the fast developments of the new detecting methods of
cancerous cells or organs in biomedical optics, such detailed theories will
contribute to have more accurate results. Finally, in the present work we
have solved the smooth double layer problem in terms of the Heun's functions.

\bigskip

\end{document}